\documentclass[aip, preprint]{revtex4-1}
\usepackage{graphicx}
\usepackage{xr}
\begin{document}

\title{Nanoscale Oxygen Defect Gradients in the Actinide Oxides}

\author{Steven R. Spurgeon}
\email{steven.spurgeon@pnnl.gov}
\affiliation{Energy and Environment Directorate, Pacific Northwest National Laboratory, Richland, Washington 99352}

\author{Michel Sassi}
\affiliation{Physical and Computational Sciences Directorate, Pacific Northwest National Laboratory, Richland, Washington 99352}

\author{Colin Ophus}
\affiliation{National Center for Electron Microscopy, Molecular Foundry, Lawrence Berkeley National Laboratory, Berkeley, California 94720}

\author{Joanne E. Stubbs}
\affiliation{Center for Advanced Radiation Sources, University of Chicago, Chicago, Illinois 60439}

\author{Eugene S. Ilton}
\affiliation{Physical and Computational Sciences Directorate, Pacific Northwest National Laboratory, Richland, Washington 99352}

\author{Edgar C. Buck}
\affiliation{Energy and Environment Directorate, Pacific Northwest National Laboratory, Richland, Washington 99352}

\date{\today}

\begin{abstract}

Oxygen defects govern the behavior of a range of materials spanning catalysis, quantum computing, and nuclear energy. Understanding and controlling these defects is particularly important for the safe use, storage, and disposal of actinide oxides in the nuclear fuel cycle, since their oxidation state influences fuel lifetimes, stability, and the contamination of groundwater. However, poorly understood nanoscale fluctuations in these systems can lead to significant deviations from bulk oxidation behavior. Here we describe the first use of aberration-corrected scanning transmission electron microscopy and electron energy loss spectroscopy to resolve changes in the local oxygen defect environment in UO$_2$ surfaces. We observe large image contrast and spectral changes that reflect the presence of sizable gradients in interstitial oxygen content at the nanoscale, which we quantify through first principles calculations and image simulations. These findings reveal an unprecedented level of excess oxygen incorporated in a complex near-surface spatial distribution, offering new insight into defect formation pathways and kinetics during UO$_2$ oxidation.

\end{abstract}

\maketitle

The engineering of oxygen defects is a central focus of modern materials science. These defects influence the electronic, magnetic, optical, and radiation-response properties of materials in ways that are difficult to control and predict \textit{a priori}.\cite{Tuller2011, Chambers2010b, Ganduglia-Pirovano2007, Sickafus2000} In particular, the safe use and disposal of oxide-based nuclear fuels depends on comprehensive models for oxidative processes and defect formation, which can guide operation, long-term waste storage, and accident cleanup efforts.\cite{Burns2012} Because of their strategic importance and potential environmental impact,\cite{Coyte2018} the oxidative behavior of the actinides has attracted considerable attention.\cite{Colmenares1984} These $5f$ elements exist in multiple valence states in oxides, leading to complex electronic properties and magnetic phase transitions that are a sensitive function of oxygen defects.\cite{Moore2009}

Among the actinides, hyper-stoichiometric UO$_{2+x}$ has been examined for over half a century because of its central role in fuel production, as well as its many interesting properties, including charge-density wave behavior and superconductivity.\cite{Cooper2018, Conradson2013, Moore2009, Shoesmith2000, McEachern1998, Fisher1994, Wasserstein1954} The system can adopt at least 14 known fluorite-derivative crystal structures, with oxidation states spanning U$^{4+} \rightarrow$ U$^{6+}$, the latter of which is aqueous soluble and a risk-driver for environmental transport.\cite{Kvashnina2013, Shoesmith2000, McEachern1998} The complex chemical landscape of this system has motivated longstanding questions about the nature of these phase transitions and the incorporation of excess oxygen. Prior work indicated that a stoichiometry of UO$_{2.25}$ is readily attainable while preserving the nominal fluorite structure;\cite{Willis1978} however, recent X-ray measurements uncovered a complex ordering of interstitial oxygen resulting from a nonclassical diffusion process.\cite{Stubbs2015} These results suggest that nanoscale deviations from predicted bulk behavior can radically change the oxidation process, calling for a more precise local understanding of the coupling between phase transitions and oxygen defect formation.

While a large body of experimental and computational work has attempted to elucidate defect formation kinetics and oxidation pathways of UO$_2$, the task is hindered by a lack of well-controlled model systems and the difficulty of simulating strongly correlated $5f$ electrons.\cite{Cooper2018, Rothe2017, Moore2009, McEachern1998} Most studies have relied on volume-averaged techniques, such as X-ray absorption near-edge structure (XANES), extended X-ray absorption fine structure (EXAFS), and X-ray photoelectron spectroscopy (XPS), applied to polycrystalline samples.\cite{Ilton2011, Caciuffo2010} When these measurements are interpreted through first principles calculations, they can yield powerful insight into electronic structure, local coordination environment, and valence.\cite{Wu1999, Jollet1997} Recent work by Stubbs \textit{et al.} has demonstrated the use of synchrotron X-ray crystal truncation rod (CTR) analysis to resolve surface distortions and subsurface oxygen interstitial profiles in single-crystal UO$_2$, albeit over a millimeter-sized region of a sample.\cite{Stubbs2017, Stubbs2015} Through fitting the CTR data and computational modeling, the authors inferred the development of oscillatory interstitial O profiles under the (001)- and (111)-oriented surfaces of UO$_2$. However, the large atomic number contrast between U and O, as well as the large lateral area over which the measurements were averaged, precluded the direct observation of interstitial geometries and localized atomic environments.

A major strength of scanning transmission electron microscopy (STEM) and electron energy loss spectroscopy (STEM-EELS) approaches is that they provide high-resolution, simultaneous information about local structure, chemistry, and defects. Past studies have shown that STEM-EELS is capable of detecting minor changes in oxidation state and composition, and that it compares favorably to X-ray results on similar uranium compounds.\cite{Jiang2018, Tobin2015, Caciuffo2010, Moore2009, Moore2004} Within the dipole approximation, these results can also be modeled using first principles methods, offering a means to quantify defect configurations and density.\cite{Jiang2018, Spurgeon2015, Aguiar2012} However, much of the pioneering STEM-EELS work on the actinides was performed several decades ago\cite{Buck2010, Colella2005, Rice1999, Xu1999, Buck1997, Fortner1997} and few studies\cite{Ochiai2018} have leveraged the advanced instrumentation or the supporting first principles computing power developed in recent years. Modern aberration-corrected microscopes, equipped with bright, sub-{\AA}ngstrom electron probes and high-speed EELS spectrometers, now permit true atomic-scale spectroscopy with exceptional energy resolution.\cite{Spurgeon2017a, Krivanek2008} Studies of complex oxides have shown that it is possible to examine image contrast\cite{Johnston-Peck2016} and spectral changes\cite{Spurgeon2017, Mundy2012d} associated with oxygen defects at interfaces and around local inhomogeneities\cite{Spurgeon2016} that lead to significant deviations from bulk properties, but are difficult to access via other means. These as-yet-untapped techniques may inform atomistic mechanisms for actinide oxidation.

Here we compare the behavior of model oxidized (001)- and unoxidized (111)-oriented single-crystal UO$_2$ surfaces using a combination of aberration-corrected STEM imaging and spectroscopy supported by first principles theory and image simulations. Previous X-ray CTR results\cite{Stubbs2017, Stubbs2015} have indicated that the former surface exhibits a two-layer periodicity in surface-normal lattice contraction, but current methods are unable to directly probe local oxygen defect configurations. We present the first atomically-resolved STEM-EELS mapping of the U $M_{4,5}$ edge, as well as a detailed examination of the O $K$ edge fine structure in the vicinity of the crystal surface. This combination of techniques provides unique insight into the basis for image contrast and the emergence of key spectral features that result from the incorporation of excess oxygen in the near-surface region. We quantify the local interstitial content at the nanoscale, finding a large amount of excess oxygen distributed across a gradient near the sample surface; however, we see no evidence for a large-scale phase transition from the fluorite structure even at stoichiometries approaching UO$_{\sim2.67}$. Finally, we identify how these gradients impact the prior understanding of UO$_2$ oxidation and discuss how they might also inform the analysis of other actinides. Our results illustrate how direct, real-space imaging approaches can reshape our understanding of oxygen defect formation in actinides with far-reaching societal impact.

\begin{figure}
\includegraphics[width=\textwidth]{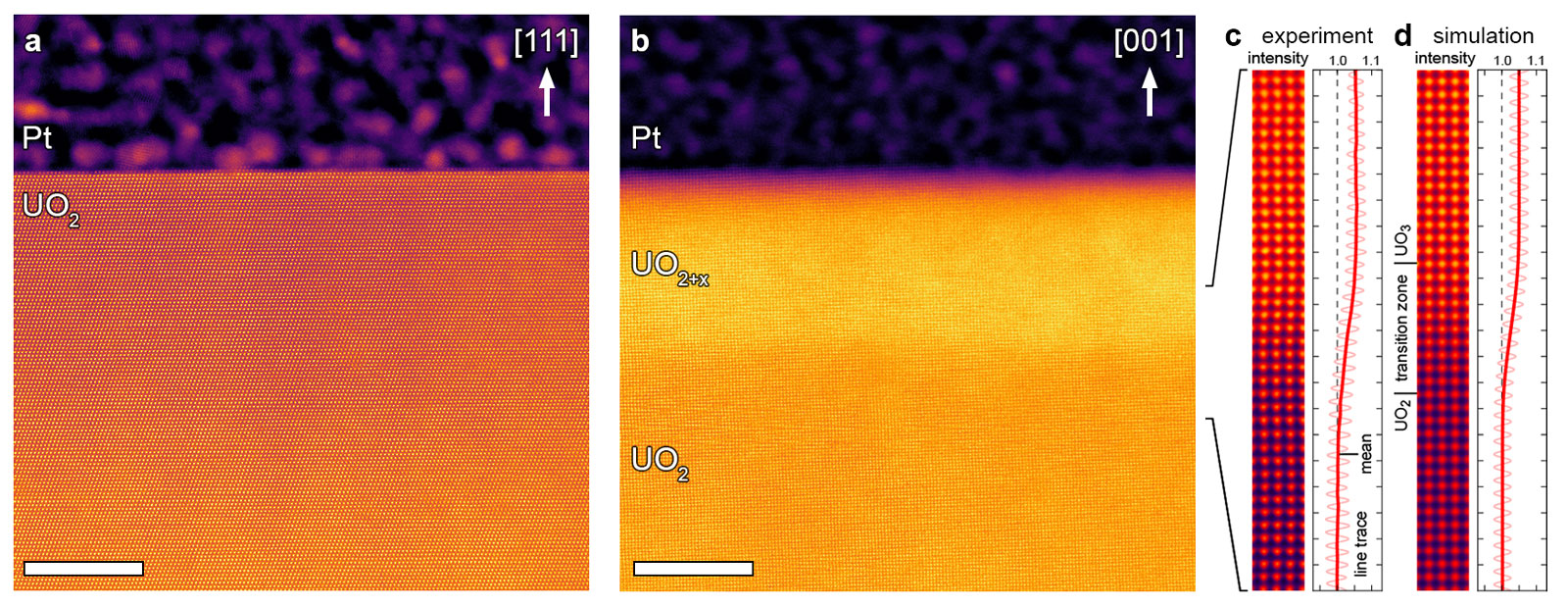}
\caption{\textbf{Imaging and simulation of the UO$_2$ sample surfaces.} (a--b) Colorized cross-sectional STEM-HAADF images of the unoxidized (111)-oriented and oxidized (001)-oriented UO$_2$ sample surfaces, respectively. These images are taken along the [110] and [100] zone-axes, respectively. Scale bars are 10 nm. (c--d) Experimental/simulated mean unit cells, and corresponding line traces, for the (001) sample. \label{haadf}}
\end{figure}

As described in the methods, we prepared two model UO$_2$ single crystal surfaces: an unoxidized (111)-oriented control sample stored in an inert gas environment and a heavily oxidized (001) sample exposed to pure O$_2$ for 21 days, followed by storage in ambient conditions for several months. The former has been predicted to be the most stable UO$_2$ surface when dry.\cite{Stubbs2017} We examined the cross-sectional structure of the near-surface region for each sample at atomic-resolution, as shown in Fig. \ref{haadf}. These images were acquired in the incoherent high-angle annular dark field (STEM-HAADF) imaging mode, whose contrast is approximately proportional to atomic number $Z^{\sim1.7}$; this mode is insensitive to the thickness-dependent contrast reversals that complicate the interpretation of typical high-resolution transmission electron micrographs.\cite{Williams2009} Both samples, shown in Figs. \ref{haadf}.a--b, exhibited a single-crystalline structure free of extended defects or impurities and we confirmed a nominal cubic fluorite structure throughout, as described in the supplementary information. However, there is a striking difference in the contrast of the (001) crystal surface, which exhibited a $\sim15$ nm band of increased intensity. This band was present in all the oxidized samples prepared and was not the result of carbon contamination or thickness variation, as confirmed by imaging and low-loss EELS shown in supplementary Figs. 1--2. We also performed geometric phase analysis (GPA) to assess possible strain variations at the nanoscale, as shown in supplementary Fig. S3. These results suggest that no large-scale phase transformation has occurred at the sample surface and that lattice bending is not responsible for the contrast band.

Interestingly, a study of CeO$_2$ nanoparticles\cite{Johnston-Peck2016} found that changes in ionic radius upon oxidation from Ce$^{3+} \rightarrow$ Ce$^{4+}$ can impart static lattice displacements to the crystal; this, in turn, can influence electron channeling and induce sizable changes in STEM image contrast. A simple estimate using Shannon ionic radii\cite{Shannon1976a} shows that a transition from eight-fold-coordinated U$^{4+}$ (1 \AA) $\rightarrow$ U$^{6+}$ (0.86 \AA) amounts to a 14\% ion size reduction, in line with the change from Ce$^{3+}$ (1.14 \AA) $\rightarrow$ Ce$^{4+}$ (0.97 \AA) of 17.5\%. While CeO$_2$ possesses a fluorite structure very similar to UO$_2$ and is often used as a proxy to simulate radiation damage effects,\cite{Jiang2017, Tracy2015} we expect even greater channeling behavior due to the higher atomic scattering factor of U ($Z = 92$) versus Ce ($Z = 58$).\cite{Peng1999} As shown in supplementary Fig. S4, we performed an array of multislice image simulations for different UO$_{2+x}$ chemistries to explore the effect of the configuration and density of oxygen defects on the resulting STEM-HAADF image contrast. Fig. \ref{haadf}.c shows an experimental mean unit cell taken from panel b, highlighting the intense contrast of the near-surface region. This experimental cell is compared to a simulation where the stoichiometry is varied from UO$_2$ to UO$_3$ over 10 lattice planes in Figs. \ref{haadf}.c--d. The simulated image intensity shows good qualitative agreement with the experimental contrast gradient of 3--5\%. A precise amount of excess interstitial oxygen cannot be determined from this comparison due to the relatively large sample thickness (55--60 nm) and computational limitations. However, our simulations combined with low loss EELS thickness measurements (see supplementary Fig. S2) indicate that the most likely explanation for the higher intensity near the surface is a large amount of interstitial oxygen, which is consistent with prior CTR analysis.\cite{Stubbs2017, Stubbs2015} These findings strongly suggest that the contrast gradient arises from changes in the local electron channeling, pointing to underlying changes in defect environment that can be probed spectroscopically.

\begin{figure}
\includegraphics[width=\textwidth]{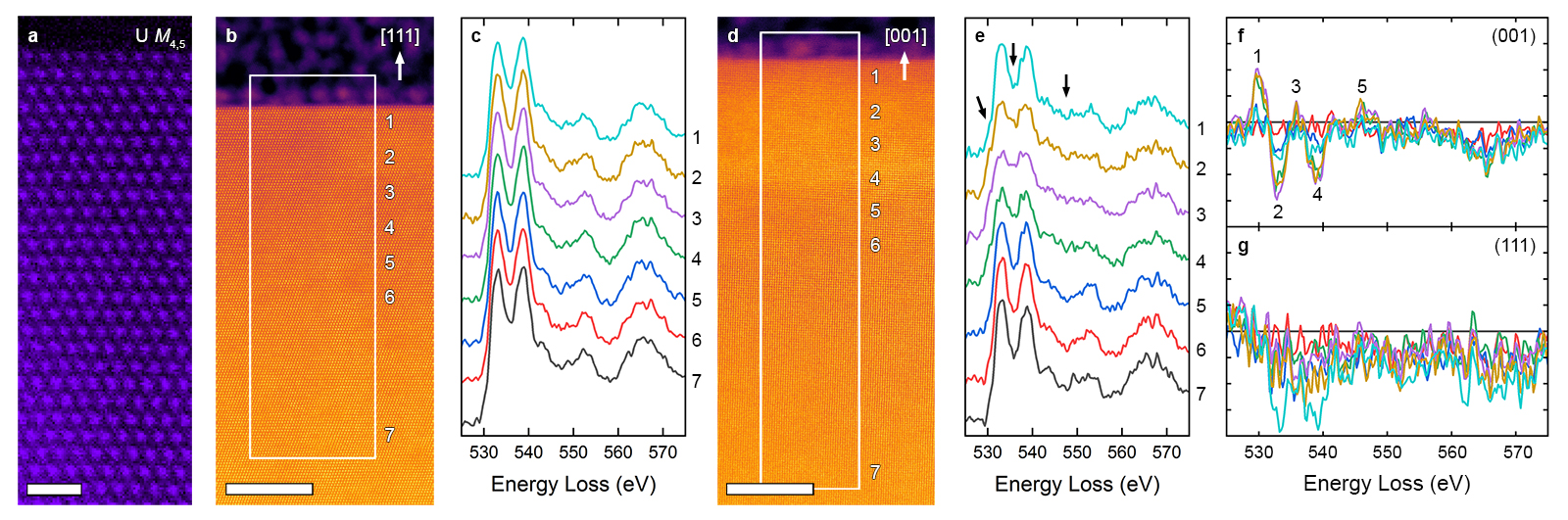}
\caption{\textbf{Spectroscopy of the UO$_2$ sample surfaces.} (a) STEM-EELS composition map of the U $M_{4,5}$ edge measured at the unoxidized (111) sample surface. Scale bar is 2 nm. (b--c) STEM-HAADF image of the (111) surface and corresponding O $K$ edge spectra extracted from the numbered regions, respectively. Scale bar is 10 nm. (d--e) STEM-HAADF image of the oxidized (001) surface and corresponding O $K$ edge spectra extracted from the numbered regions, respectively. Scale bar is 10 nm. (f--g) Difference plots showing changes in spectral features relative to the bulk, marked by \textbf{1}--\textbf{5}, for the (001)- and (111)-oriented sample surfaces, respectively.\label{eels}}
\end{figure}

Accordingly, we performed atomic-scale STEM-EELS mapping of the U $M_{4,5}$ and O $K$ ionization edges for the two samples, as shown in Fig. \ref{eels}. The U $M_{4,5}$ edge results from white-line transitions from the U $3d \rightarrow 5f$ states;\cite{Degueldre2013} its higher ionization energy compared to the $N_{4,5}$ ($4d \rightarrow 5f$) and $O_{4,5}$ ($5d \rightarrow 5f$) edges makes it an excellent candidate for localized composition mapping. The O $K$ edge results from transitions from the O $2p \rightarrow$ U $5f$ and $6d$ states;\cite{Wu1999,Jollet1997} while more complex, this edge encodes detailed information about the U coordination environment and therefore offers a window into defects formed during oxidation. Although previous studies have examined trends in the U $M_{4,5}$ white-line ratio in EELS,\cite{Colella2005, Fortner1997} hardware limitations made it difficult to perform atomic-resolution mapping needed to understand local fluctuations in composition and chemistry at surfaces and interfaces. With the development of modern aberration correctors, the large probe convergence angles, high currents, and small probe sizes needed for atomic-scale spectroscopy are now available.

We focused our spectroscopic mapping on the problem of identifying signatures of interstitial oxygen defects, which required supporting first principles calculations. Fig. \ref{eels}.a shows a composition map of the U $M_{4,5}$ edge collected at the unoxidized (111) surface, illustrating the excellent compositional uniformity and crystallinity of the sample up to its top monolayer. Figs. \ref{eels}.b--c show O $K$ edge spectra extracted from a map of the (111) surface region, progressing from the topmost layer of the crystal to its bulk. The edge displays two sharp peaks at 533 and 538.75 eV, followed by a shoulder at 543.25 eV and two broader features at 553 and 566 eV. We note that the overall line shape is in good agreement with prior work\cite{Aguiar2012} and that there is little variation upon moving from the surface (spectrum 1) to the bulk (spectrum 7).

In comparison, O $K$ edge measurements of the oxidized (001) surface, shown in Figs. \ref{eels}.d--e, exhibit markedly different behavior. The overall bulk spectrum 7 is quite similar to the (111) sample, with a minor change in the ratio of the two sharp peaks at lower energy loss. However, moving closer to the sample surface, there is a striking redistribution of spectral features that coincides with the presence of the surface contrast band (spectra 1--4). These features are highlighted in the difference spectra in Figs. \ref{eels}.f--g. Notably, we observe the emergence of a distinct shoulder at 530.5 eV (feature \textbf{1}), changes in the ratio of the two main peaks (features \textbf{2} and \textbf{4}), an increase in the minimum at  535.5 eV (feature \textbf{3}), as well as the emergence of a broad peak at 548 eV (feature \textbf{5}). Collectively these results suggest the possibility of distinct local structures giving rise to these spectral features in the oxidized (001) surface.

\begin{figure}
\includegraphics[width=\textwidth]{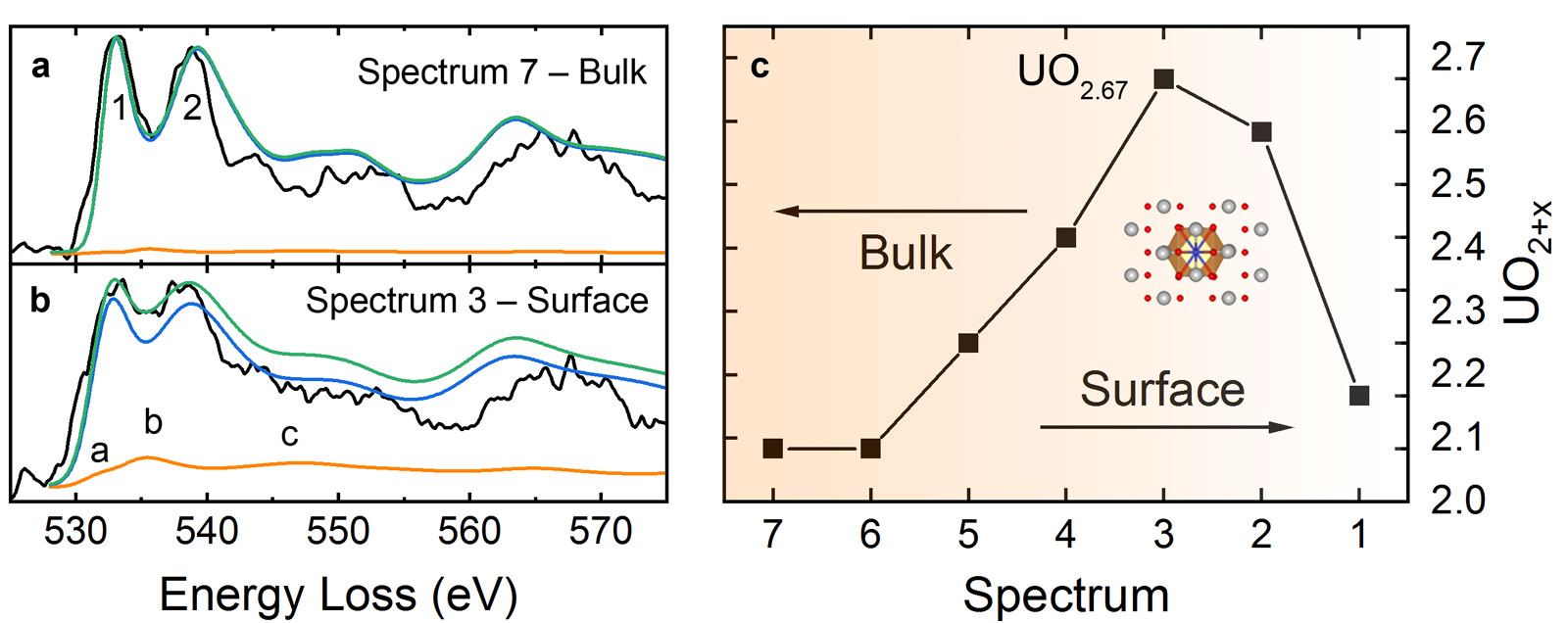}
\caption{\textbf{Analysis of oxygen defect environment in the (001) sample.} (a--b) Comparison between experimental EELS and theoretically calculated XANES spectra accounting for the presence of interstitial oxygen for the bulk spectrum 7 and near-surface spectrum 3, respectively. The experimental data (black), contributions from O lattice sites (blue), interstitial O (orange), and the resulting linear combinations (green) are shown. Key emergent features are marked by \textbf{a}--\textbf{c}. (c) Estimate of effective local stoichiometry per unit cell. The inset shows an illustration of the oxygen defect formed.\label{dft}}
\end{figure}

We therefore turned to first principles theory calculations using the density functional theory (DFT) framework to explore possible defect structures that could give rise to the measured spectral features. Within the dipole approximation, calculated XANES spectra can be compared to measured EELS spectra, since they probe the same electronic states; these comparisons provide valuable insight and are commonly used to rationalize observed trends and fine structure features in oxides.\cite{Spurgeon2015} While a ground state approach cannot account for excited core-hole effects, they have been shown to have little influence on the predicted energy loss near-edge structure (ELNES) in the fluorites.\cite{Aguiar2010} Using the FDMNES code,\cite{Bunau2009} we calculated O $K$ edge XANES spectra for both pristine and defective UO$_2$ containing either an oxygen interstitial or vacancy (see supplementary information).

The calculated spectrum for bulk defect-free UO$_2$ is in good agreement with experimental measurements deep in the bulk of the sample; as shown in supplementary Fig. S6, all the spectral features and relative intensity of the two sharp experimental peaks at 533 and 538.75 eV are well reproduced. The introduction of an interstitial O in the lattice deforms the surrounding environment, breaking the high cubic symmetry of UO$_2$ and inducing a degeneracy in U--O bonding. Accordingly, two component spectra were calculated: one is the average of all the lattice site O atoms, and the other is for the interstitial O atom, shown by the blue and orange curves in Fig. \ref{dft}, respectively. These two component spectra show differences compared to the one calculated for a defect-free high cubic symmetry UO$_2$ (see supplementary Fig. S7); in particular, for the lattice O sites we find that the relative intensity of the two first peaks, labelled \textbf{1} and \textbf{2} in Fig. \ref{dft}.a, is reversed, with peak \textbf{2} being less intense than peak \textbf{1}. Especially interesting is the calculated spectrum for the interstitial O atom in UO$_2$, which exhibits completely different spectral features compared to the one from lattice O sites. Three features---labeled \textbf{a}, \textbf{b}, and \textbf{c}---emerge that correspond to the positions of the experimentally observed changes, as shown in Fig. \ref{dft}.b. These features are positioned such that the two minima of the interstitial O spectra, located between \textbf{a} and \textbf{b}, and \textbf{b} and \textbf{c}, overlap with peaks \textbf{1} and \textbf{2} of the lattice O site spectra, while the peak \textbf{b} is aligned with the minimum in-between peak \textbf{1} and \textbf{2}. Another interesting characteristic in the spectra of interstitial O is that the minimum between \textbf{b} and \textbf{c} has a higher intensity than the minimum between \textbf{a} and \textbf{b}. Therefore, performing a linear combination of the calculated lattice and interstitial O $K$ edge component spectra leads to an increase of the intensity of peak \textbf{2} greater than that of peak \textbf{1}, and fills the minimum between those peaks, as seen by the green linear combination fit curves shown in Fig. \ref{dft}.

To compare and qualitatively reproduce the changes observed in the oxidized (001) sample, we varied the contribution of the O interstitial component in the linear combination and varied the broadening of the calculated spectra to best match the experimental one for each case. As shown in Figs. \ref{dft}.a--b, the best fit---considering the relative intensity between peaks \textbf{1} and \textbf{2}, as well as the minimum in-between---yields a much larger amount of interstitials for the near-surface (spectrum 3) than for the bulk (spectrum 7). The fits suggest that the stoichiometry of the material in the region of spectrum 3 is closer to UO$_{2.6667}$, while it is less than UO$_{2.0833}$ for the bulk region of spectrum 7 (nominally UO$_2$), as shown in Fig. \ref{dft}.c. Recognizing that the overall trend is certainly more reliable than absolute compositional values, it is nonetheless clear that there is a gradient in the interstitial distribution in the (001) sample and that the region near spectra 3 contains significantly more interstitial O atoms compared to the bulk region.

These findings are consistent with our imaging and multislice simulations, which show that a sizable interstitial O content is needed to reproduce the experimental contrast. While a combination of other defects may be involved in the change of spectral features, we note that neither O vacancies nor a homogeneous lattice expansion or contraction can effectively reproduce the spectral changes. As shown in supplementary Fig. S7, the filling of the minimum in-between peaks \textbf{1} and \textbf{2}, as well as the overall intensity increase of the post-peak \textbf{2} shoulder, cannot be reproduced in these scenarios. Rather, we find that the unique spectral features observed in the sample can best be described by interstitial oxygen defects that affect the U coordination environment.

In summary, we observe an unprecedented large amount of interstitial oxygen distributed across a nanoscale gradient in the (001) surface. Importantly, we find no evidence for large-scale phase transformations, suggesting that a stoichiometry of nearly UO$_{2.67}$ is attained in the fluorite structure; this value is far in excess of the UO$_{2.25}$ expected from prior reports and represents a significant departure from bulk behavior. Considering the non-equilibrium nature of the oxidative process, we emphasize the competition of the bulk and surface states of the crystal. Analogous to substrate-induced ``clamping'' in multiferroic oxide heterostructures,\cite{Bichurin2003} it is likely that structural distortions of the surface are constrained by the underlying bulk, limiting associated phase transitions. The competition between transport of oxygen into the bulk and reduction of the surface may also give rise to the observed gradient in oxygen interstitial content. These new nanoscale insights can help refine our understanding of oxygen transport and defect formation kinetics in this system. More broadly, our study shows how a combination of STEM, EELS, and first principles calculations may be used to fingerprint the local chemical environment of actinide surfaces. Substantial excess oxygen manifests in image contrast changes that are accompanied by unique spectral signatures in the O $K$ edge fine structure. Our simulations are able to disentangle key contributors to these signatures, identifying likely oxygen configurations that can be examined in other related systems. Using this approach, it is now possible to examine defect generation at surfaces and interfaces, such as grain boundaries in fuel assemblies and waste forms, paving the way for more comprehensive atomistic models for oxidation of the actinides.

\clearpage

\section*{Acknowledgements}

S.R.S. thanks Drs. Kevin Rosso, Demie Kepaptsoglou, and Lewys Jones for helpful discussions. Pacific Northwest National Laboratory (PNNL) is a multiprogram national laboratory operated for DOE by Battelle.  This work was supported by the Nuclear Process Science Initiative (NPSI) at PNNL. A portion of the microscopy was performed in the Radiological Microscopy Suite (RMS), located in the Radiochemical Processing Laboratory (RPL) at PNNL. Work at the Molecular Foundry was supported by the Office of Science, Office of Basic Energy Sciences, of the U.S. Department of Energy under Contract No. DE-AC02-05CH11231. C.O. acknowledges support from the DOE Early Career Research Program. J.E.S. is supported by the National Science Foundation - Earth Sciences (EAR - 1634415) and Department of Energy- GeoSciences (DE-FG02-94ER14466 and DE-SC0019108). We thank R. Caciuffo (Institute for Transuranium Elements) and M. Paffett (Los Alamos National Laboratory) for providing the UO$_2$ crystals.

\section*{Author Contributions}

S.R.S., M.S., C.O., E.S.I., and E.B. planned the experiments. S.R.S. conducted imaging and EELS analysis. M.S. performed first principles calculations. C.O. performed multislice simulations. J.E.S. prepared the samples. All authors contributed to the data interpretation and manuscript preparation.

\section*{Additional Information}

Supplementary information detailing the methods used, imaging, thickness measurements, strain analysis, multislice simulation, and first principles calculations is available in the online version of the paper. Reprints and permissions information is available online at www.nature.com/reprints. Correspondence and requests for materials should be addressed to S.R.S.

\section*{Competing Financial Interests}

The authors declare no competing financial interests.

\clearpage

\bibliography{References}

\end{document}

\section*{Supplementary Note 1: Methods}

\subsection*{Sample Preparation}

We have prepared polished single crystals of UO$_2$ oriented within 0.1$^{\circ}$ of the (001) and (111) surface planes, as described elsewhere.\cite{Stubbs2015} The (111) sample was an unoxidized control. The (001) sample was exposed to 1 atm of dry oxygen gas for 21 days and measured on the GSECARS Beamlines 13-IDC and 13-BMC at the Advanced Photon Source, Argonne National Laboratory.\cite{Stubbs2017} It was subsequently stored in air for several months prior to STEM preparation.

\subsection*{STEM Imaging}

Cross-sectional STEM samples were prepared using an FEI Helios NanoLab DualBeam Focused Ion Beam (FIB) microscope and a standard lift out procedure along the UO$_2$ [100] and [110] zone-axes for the (001)-oriented and (111)-oriented samples, respectively. Initial cuts were made at 30 kV / 1.5$^{\circ}$ and final polishing at 2 kV / 2.5$^{\circ}$ ion beam energy / incidence angles. Images and fine structure maps were collected on a probe-corrected JEOL ARM-200CF microscope operating at 200 kV accelerating voltage, with a probe semi-convergence angle of 27.5 mrad, a HAADF collection angle of 82--186 mrad, and a EELS inner collection angle of 42.9 mrad. EELS fine structure maps were collected using a 1~\AA~probe size with a $\sim130$ pA probe current and a 0.25 eV ch$^{-1}$ dispersion, yielding an effective energy resolution of $\sim0.75$~eV. The O $K$ edge spectra were corrected for energy drift using the zero-loss peak, then treated with a power low background subtraction fit to a 60 eV window prior to the edge, and processed using a Fourier-Ratio deconvolution. Separate composition maps were collected on a probe-corrected JEOL ARM-300F microscope operating at 300 kV accelerating voltage, with a probe semi-convergence angle of 28 mrad and a EELS inner collection angle of 87 mrad with a 1~\AA~probe size with a $\sim260$ pA probe current and a 1 eV ch$^{-1}$ dispersion. To improve signal-to-noise, the spectrometer was binned 4$\times$ in the energy axis.

\subsection*{Multislice Simulations}

We have performed STEM image simulations to determine the effect of interstitial and total oxygen on the STEM-HAADF image intensity for UO$_2$ along the [100] zone-axis. These image simulations were performed using the PRISM method described in Ref. \onlinecite{Ophus2017}, implemented in the Prismatic software described in Ref. \onlinecite{Pryor2017}. An accelerating voltage of 200 kV, a probe semi-convergence angle of 27.5 mrad, and HAADF collection angles of 82--186 mrad were set to match the experimental parameters. An in-plane pixel sampling of 0.0337~\AA~and a slice thickness of 1.3678~\AA~were used. A total thickness of 60 nm was used for all simulations, with an in-plane tiling of $5 \times 5$ UO$_2$ unit cells to minimize probe wrap-around errors. 25 frozen phonon configurations were used to include thermal scattering effects. A PRISM interpolation factor of 1 was used in the $x$ and $y$ directions, making the simulation mathematically identical to the multislice method.\cite{Ophus2017} The atomic scattering potentials used and more information about the multislice method are given in Ref. \onlinecite{Kirkland2010}. The oxygen concentration was independently varied on two sublattice sites---the bulk structure and interstitial sites---for total stoichiometries from UO$_2$ to UO$_3$, shown in Fig. \ref{si_multislice_table}.

\subsection*{Density Functional Theory Calculations}

In this study, two set of computational simulations were performed. First, the defect-free and defective crystal structures of UO$_2$ were optimized in the density functional theory (DFT) framework, as implemented in the VASP package,\cite{Kresse1996a, Kresse1996b} with the generalized gradient approximation (GGA) and the Perdew-Burke-Ernzerhof (PBE) parametrization\cite{Perdew1996} exchange-correlation functional. In each calculation, the cutoff energy of the projector augmented wave\cite{Blochl1994} pseudo-potential was 600 eV and a Gamma centered $k$-points mesh of $6\times6\times6$ for the sampling of the Brillouin zone was used. The total energy was converged to 10$^{-5}$ eV cell$^{-1}$ and the force components were relaxed to below 10$^{-3}$ eV \AA$^{-1}$. Spin-polarization and the Vosko-Wilk-Nusair local density approximation scheme\cite{Vosko1980} were used. The GGA+U method, as described by Dudarev,\cite{Dudarev1998} was used for the U atoms to correct the description of the Coulomb repulsion of the 5$f$ electrons in standard GGA. The Hubbard parameter, $U$, describing the Coulomb interaction, was fixed to 4.5 eV, while the screened exchange energy, $J$, was fixed to 0.51 eV.\cite{Dorado2009} Subsequently, we used the relaxed defect-free and defective supercell to calculate the O $K$ edge XANES spectra with the FDMNES code.\cite{Bunau2009} Although EELS and XANES are not strictly equivalent techniques, they probe the same electronic states. Therefore, a comparison between experimental and theoretical spectra across the two techniques provide invaluable insight and are commonly used to rationalize observed trends and fine structure features in oxides.\cite{Spurgeon2015} In FDMNES, the final excited state is obtained by solving a Schr{\"o}dinger-like equation through the Green’s formalism, within the limit of the muffin-tin approximation. The potentials and Fermi energy were determined self-consistently using a radii of 7 \AA. Similar radii were used for the calculations of the spectra. Real Hedin-Lundquist potentials\cite{Hedin1971} were used to model the exchange-correlation. Dipoles, quadrupoles, core-hole and spin-orbit contributions were taken into account. All the DFT calculations used a $2 \times 2$ (111) hexagonal unit cell of UO$_2$ containing $12\times$ U and $24\times$ O atoms. The position of the interstitial O atom in this unit cell was identical to the one proposed in Ref. \onlinecite{Stubbs2015} and is shown in Fig. \ref{si_int_model}.

\clearpage

\section*{Supplementary Note 2: Uniformity of Surface Oxidation}

We have measured multiple regions of the oxidized (001) sample and confirm the presence of a uniform, high-contrast band at the surface. Moreover, we observe minimal long-range defects and impurities, reflecting the high quality of the sample.

\begin{figure}[h]
\includegraphics[width=\textwidth]{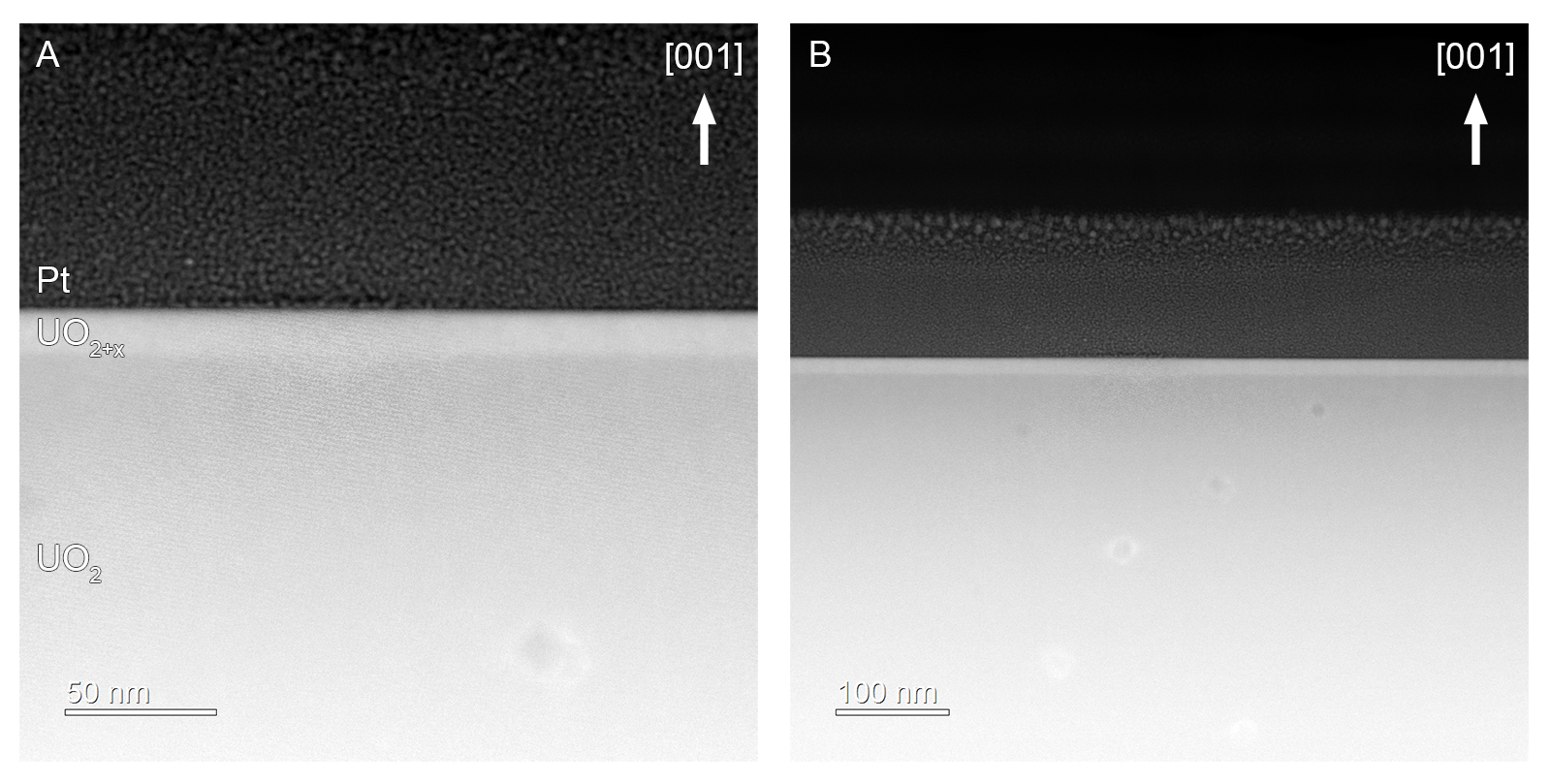}
\caption{(A) Representative intermediate- and (B) low-magnification cross-sectional STEM-HAADF images of the (001) sample. \label{si_overview}}
\end{figure}

\clearpage

\section*{Supplementary Note 3: Electron Energy Loss Spectroscopy Thickness Mapping}

To assess the possibility of local thickness variations, we have performed EELS thickness mapping of the surface band in the (001) sample. We find that the sample varies from 55--60 nm and that no abrupt change in thickness occurs that might indicate carbon contamination or lattice bending.

\begin{figure}[h]
\includegraphics[width=\textwidth]{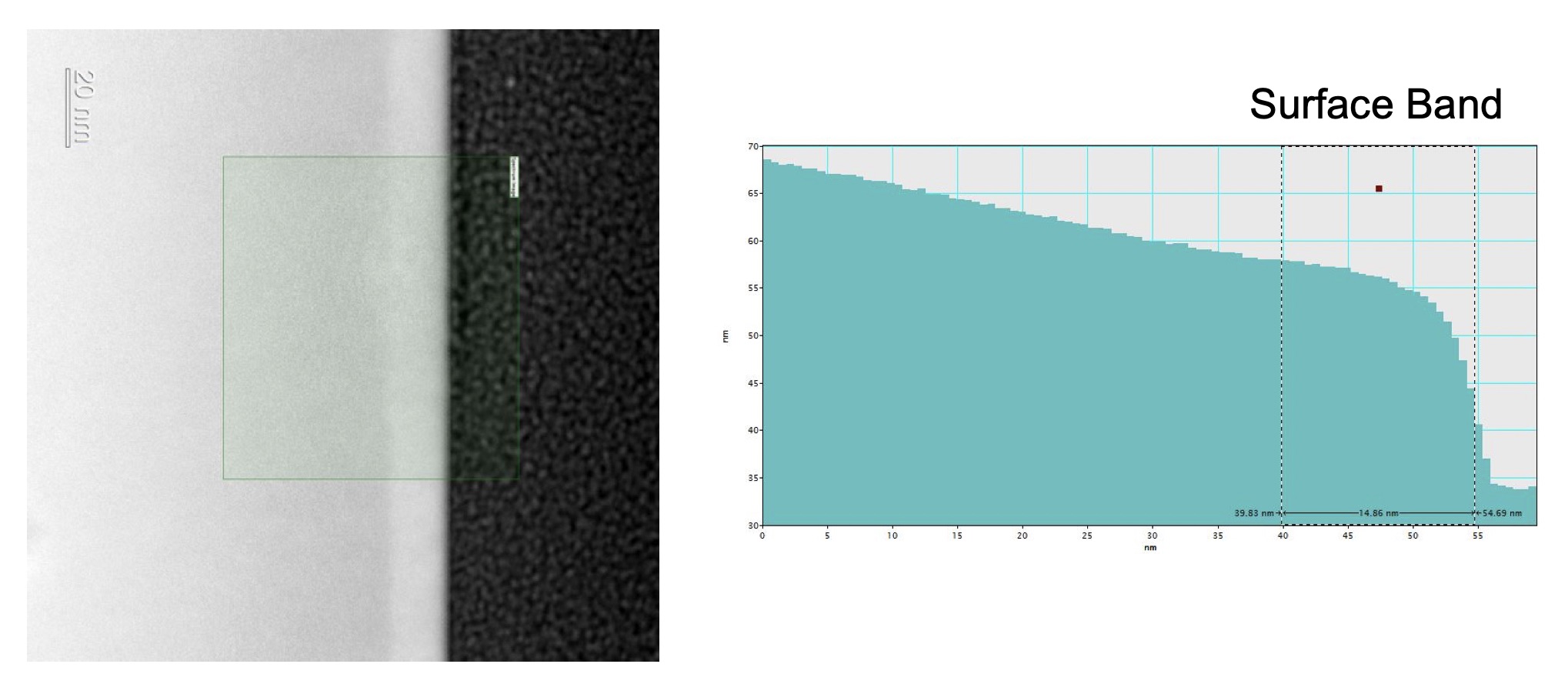}
\caption{(Left) STEM-HAADF image of the (001) sample and (Right) associated STEM-EELS thickness map. \label{si_eels_thickness}}
\end{figure}

\clearpage

\section*{Supplementary Note 4: Geometric Phase Analysis Strain Mapping}

We confirm the nominal UO$_2$ fluorite crystal structure for both samples shown in the main text. The (111) sample has experimentally measured reciprocal lattice vectors of $\textbf{g}_{[1\bar{1}\bar{1}]} = 0.314$ \AA$^{-1}$, $\textbf{g}_{[00\bar{2}]} = 0.362$ \AA$^{-1}$, $\textbf{g}_{[\bar{1}1\bar{1}]} = 0.514$ \AA$^{-1}$ versus expected bulk values of $\textbf{g}_{[1\bar{1}\bar{1}]} = 0.318$ \AA$^{-1}$, $\textbf{g}_{[00\bar{2}]} = 0.367$ \AA$^{-1}$, $\textbf{g}_{[\bar{1}1\bar{1}]} = 0.519$ \AA$^{-1}$.\cite{Wasserstein1954} The (001) sample has experimentally measured reciprocal lattice vectors of $\textbf{g}_{[200]} = 0.363$ \AA$^{-1}$, $\textbf{g}_{[0\bar{2}0]} = 0.362$ \AA$^{-1}$, $\textbf{g}_{[2\bar{2}0]} = 0.514$ \AA$^{-1}$ versus expected bulk values of $\textbf{g}_{[200]} = 0.367$ \AA$^{-1}$, $\textbf{g}_{[0\bar{2}0]} = 0.367$ \AA$^{-1}$, $\textbf{g}_{[2\bar{2}0]} = 0.519$ \AA$^{-1}$.

\begin{figure}[h]
\includegraphics[width=\textwidth]{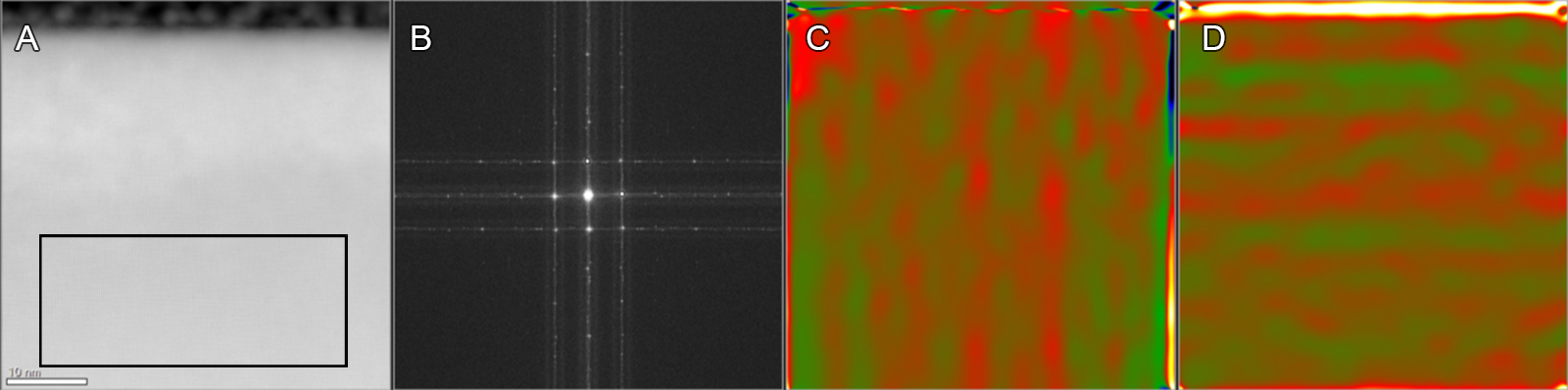}
\caption{(A) Rigid-aligned STEM-HAADF image, (B) Corresponding diffraction pattern, (C) In-plane strain component ($\epsilon_{xx}$) and (D) Out-of-plane strain component ($\epsilon_{yy}$). The rectangle indicates the reference lattice. \label{si_gpa}}
\end{figure}

To explore the local variation in strain state, we have performed geometric phase analysis (GPA) using the FRWRtools plugin developed by Cristoph Koch.\cite{Koch2002} In theory, GPA is capable of measuring lattice strains at the sub-nanometer length scale, with better than 0.1\% strain resolution.\cite{Hytch2014, Hytch2007} However, the technique is prone to stripe-like artifacts resulting from scan distortions in STEM,\cite{Vatanparast2017} so we have collected a series of drift-corrected images using the SmartAlign plugin, which are subsequently rigid-aligned and averaged.\cite{Jones2015} Fig. \ref{si_gpa} shows the imaged region, diffraction pattern, and corresponding in- and out-of-plane strain maps. While some random variations are present in the data, we observe no clear and systematic lattice distortion in the surface region that might indicate lattice bending or a large-scale phase transformation. This result is not surprising, considering the small expected strains and possible local variations associated with oxygen interstitial incorporation.

\clearpage

\section*{Supplementary Note 5: Multislice Simulations of Interstitial Oxygen}

\begin{figure}[h]
\includegraphics[width=\textwidth]{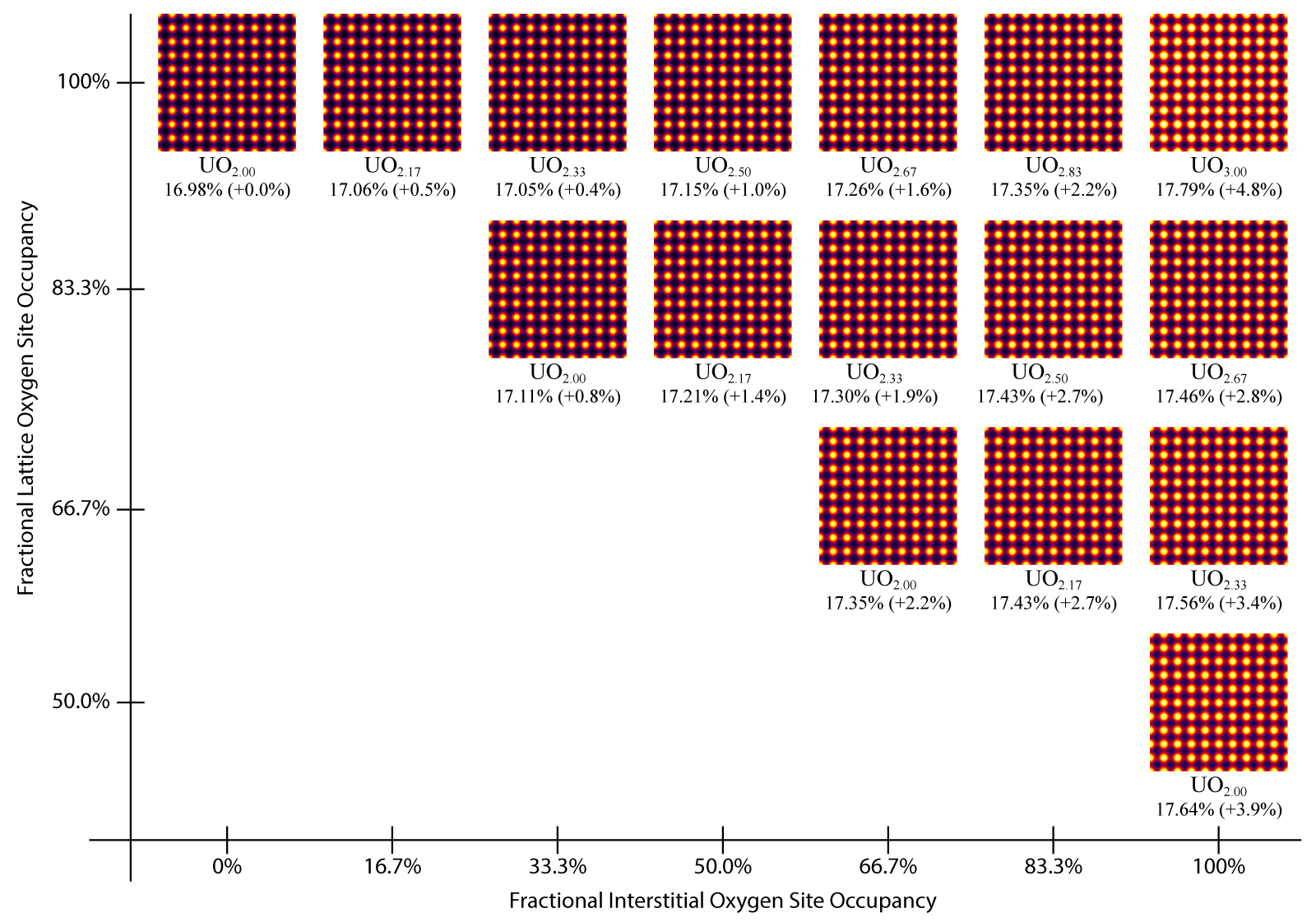}
\caption{Table of image simulations for different lattice and fractional interstitial oxygen site occupancies. \label{si_multislice_table}}
\end{figure}

\begin{figure}
\includegraphics[width=0.6\textwidth]{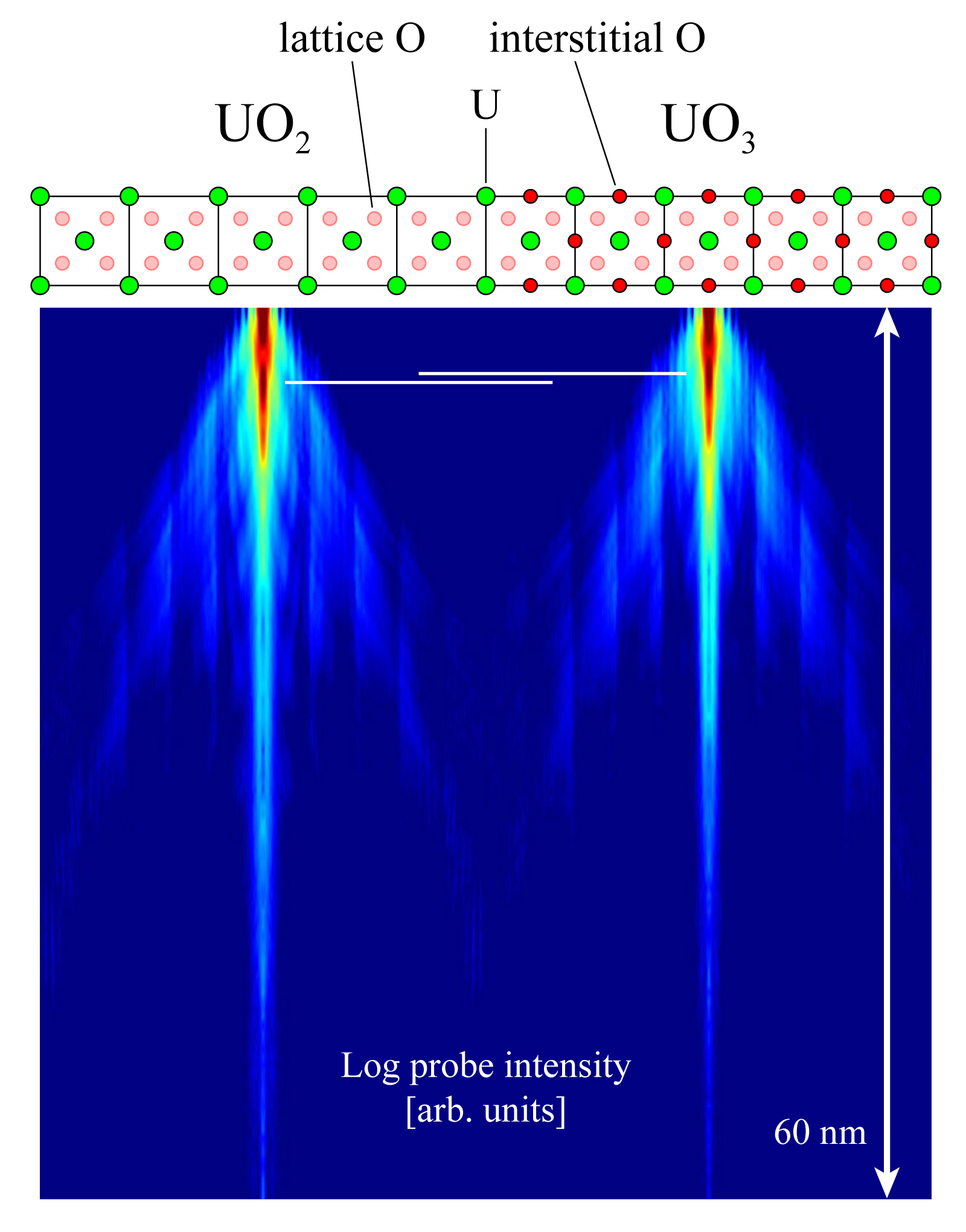}
\caption{Calculated differences in electron probe scattering for UO$_2$ and UO$_3$. \label{si_multislice_channeling}}
\end{figure}

We have performed an array of multislice simulations for a range of different fractional lattice and oxygen site occupancies, as shown in Fig. \ref{si_multislice_table}. We find that there are pronounced differences in image contrast that can be accounted for primarily by the increased scattering cross section caused by excess interstitial oxygen. The mean intensity of each STEM-HAADF image is given as percentage of the total probe current, and the relative percentage change from the bulk UO$_2$ signal is shown in brackets. Excess oxygen in the interstitial sites also shortens the channeling length along the U + interstitial O columns, as shown in Fig. \ref{si_multislice_channeling}. Additional channeling contrast along adjacent atomic columns is also visible. These simulations confirm that the observed experimental contrast change can be achieved through the incorporation of excessive interstitial oxygen.

\clearpage

\section*{Supplementary Note 6: Details of Density Functional Theory Calculations}

\begin{figure}[h]
\includegraphics[width=0.5\textwidth]{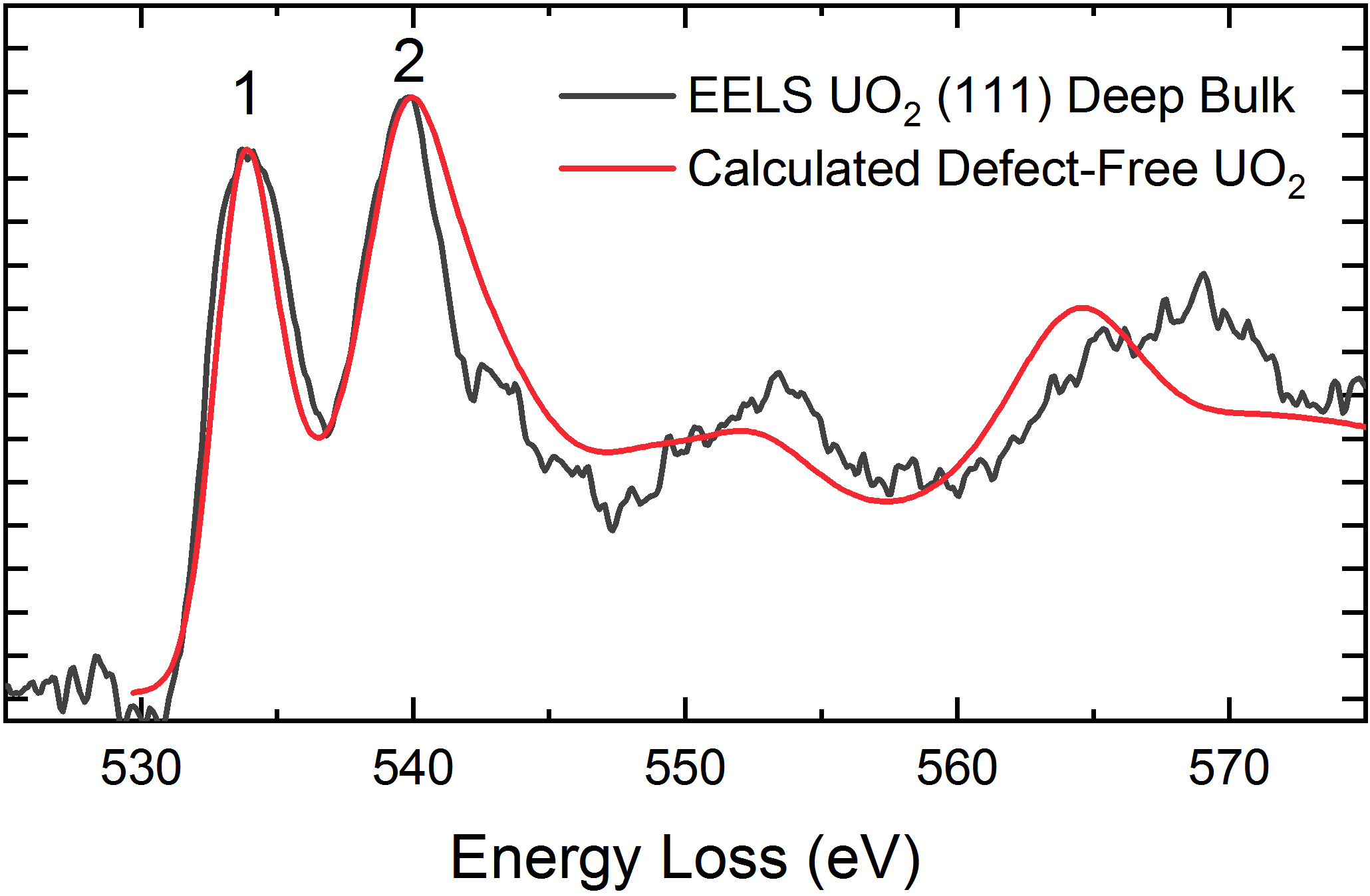}
\caption{Comparison between experimental O $K$ edge spectrum of UO$_2$ taken deep in the bulk ($\sim$100 nm from surface) of the (111) sample and the calculated spectrum for a defect-free UO$_2$ lattice. \label{si_dft_bulk}}
\end{figure}

\begin{figure}[h]
\includegraphics[width=0.5\textwidth]{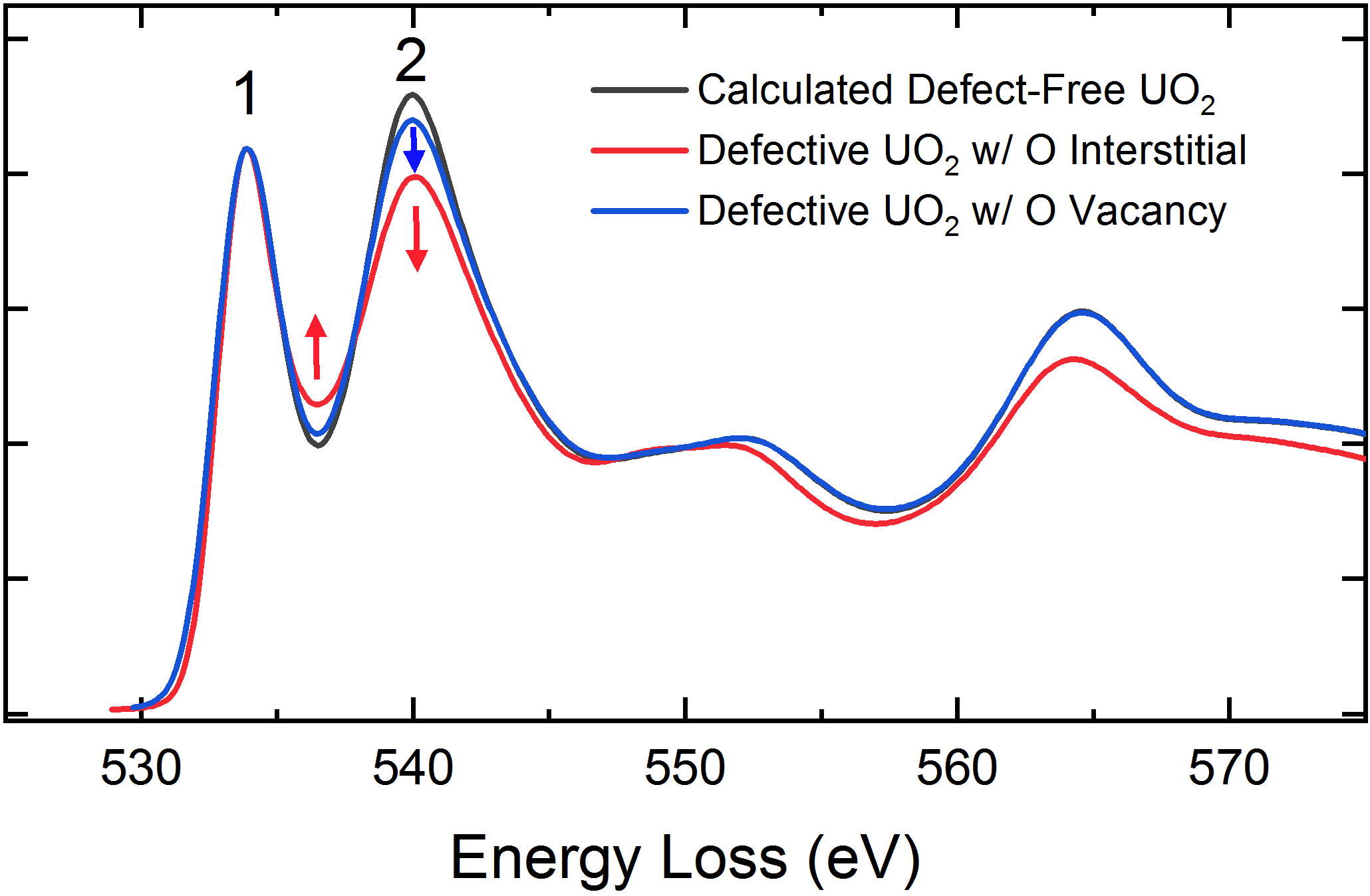}
\caption{Comparison of the spectral changes associated with an oxygen interstitial and vacancy. \label{si_dft_vac}}
\end{figure}

\begin{figure}[h]
\includegraphics[width=0.5\textwidth]{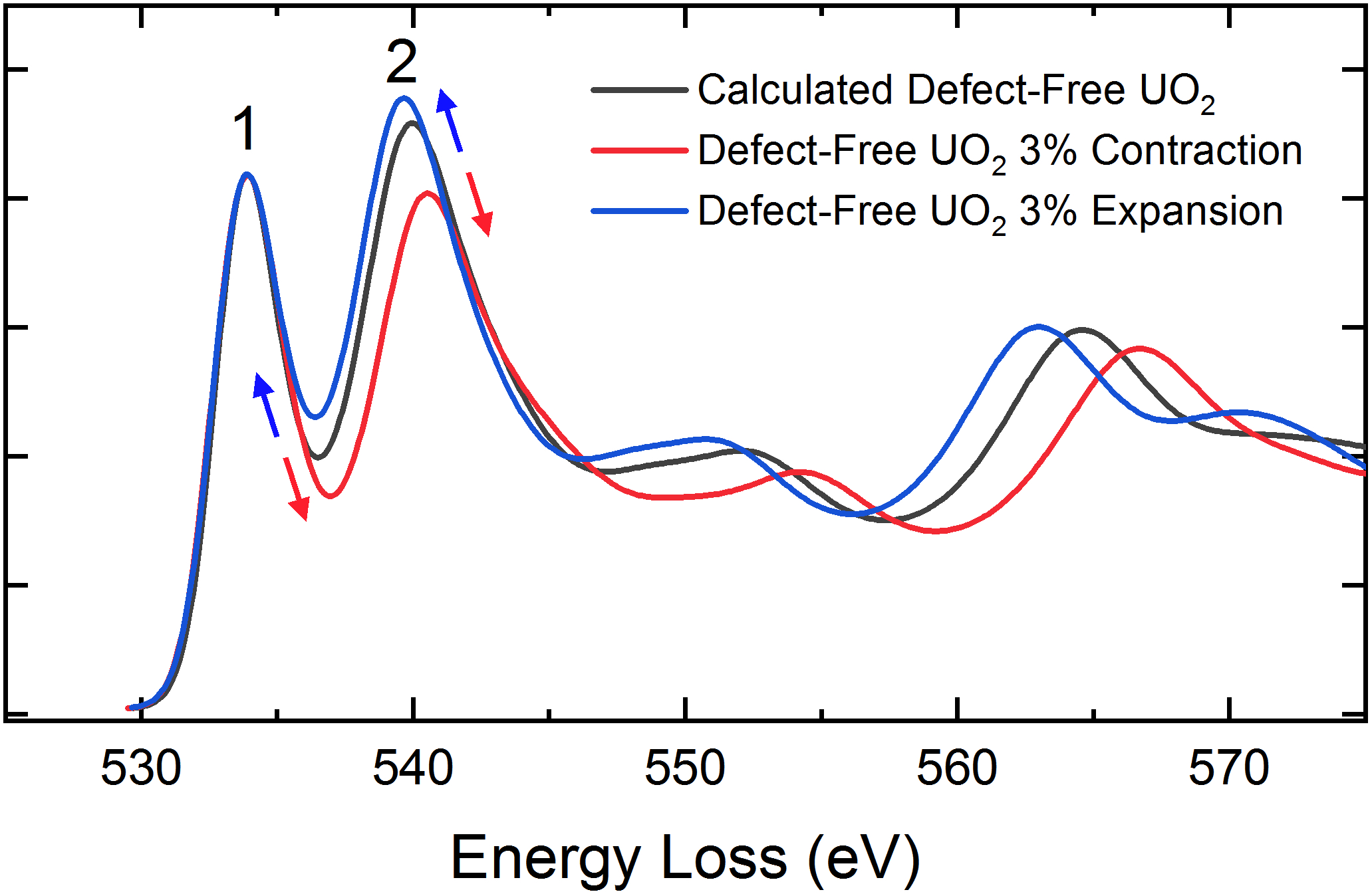}
\caption{Comparison of the spectral changes associated with homogeneous lattice expansion and contraction of $a$, $b$, and $c$ lattice parameters of cubic UO$_2$. \label{si_dft_exp_cont}}
\end{figure}

\begin{figure}[h]
\includegraphics[width=0.7\textwidth]{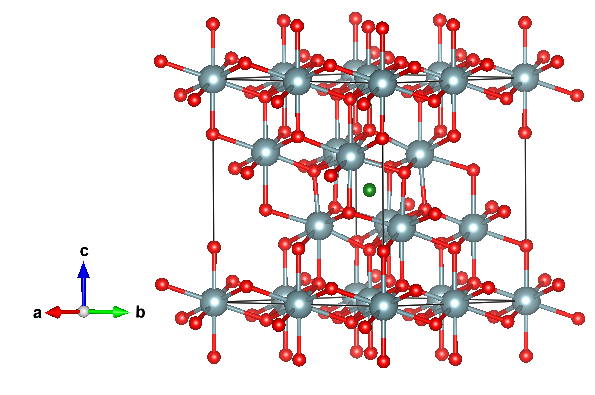}
\caption{Representation of interstitial O in UO$_2$ after relaxation by DFT calculations. \label{si_int_model}}
\end{figure}

Figure \ref{si_dft_bulk} shows a comparison between the calculated and measured UO$_2$ EELS spectra measured deep in the bulk of the unoxidized (111) sample. Figure \ref{si_dft_vac} compares the spectral modifications induced by O interstitials and vacancies. While O vacancies can reduce the intensity of peak \textbf{2}, they affect the intensity of the minimum in between peak \textbf{1} and \textbf{2} to a lesser degree than interstitial O, and have no noticeable effects on the post-peak \textbf{2} shoulder. The effects of uniform lattice expansion and contraction are anti-correlated, as shown in Fig. \ref{si_dft_exp_cont}. While lattice contraction can reduce the relative intensity of peak \textbf{2} with respect to peak \textbf{1}, it does not fill the minimum in between those peaks. Finally, Figure \ref{si_int_model} shows the particular lattice and defect representation used in our calculations.

\clearpage

\bibliography{References}